\documentclass[epj,draft]{svjour}
\usepackage{amsmath}
\usepackage{cite}
\usepackage{graphics}
\usepackage{epsf}
\usepackage{multirow}
\usepackage{dcolumn}
\usepackage{url}
\usepackage{subfigure}
\begin{document}
\title{Melting tungsten nanoparticles: a molecular
  dynamics study}


\author{Amitava~Moitra\inst{1,2} \and Sungho~Kim\inst{1,2} %
\and J.~Houze\inst{1,2} \and B.~Jelinek\inst{1,2} %
\and Seong-Jin~Park\inst{2} \and Randall~M.~German\inst{3,2} %
\and M.~F.~Horstemeyer\inst{2} %
\and Seong-Gon~Kim\inst{3,2} %
}                     
%
\mail{Seong-Gon~Kim, \url{kimsg@ccs.msstate.edu}}
\institute{%
Department of Physics and Astronomy, 
Mississippi State University,
Mississippi State, MS 39762, USA
\and %
Center for Advanced Vehicular Systems, 
Mississippi State University, 
Mississippi State, MS 39762, USA
\and %
Department of Mechanical Engineering,
Mississippi State University,
Mississippi State, MS 39762, USA
}
\date{Received: date / Revised version: date}
%
\abstract{
  We report a molecular dynamics simulation of melting of tungsten (W)
  nanoparticles.  The modified embedded atom method (MEAM) interatomic
  potentials are used to describe the interaction between tungsten
  atoms.  The melting temperature of unsupported tungsten
  nanoparticles of different sizes are found to decrease as the size
  of the particles decreases.  The melting temperature obtained in the
  present study is approximately a decreasing function of inverse
  radius, in a good agreement with the predictions of thermodynamic
  models.  We also observed that the melting of a W nanoparticle is
  preceded by the premelting of its outer skin at a temperature lower
  than its melting temperature.
} 
\maketitle
\section{Introduction}
\label{intro}
Tungsten, along with its alloys and compounds, occupies a unique
position in materials science. The material properties that make
tungsten attractive to the metals industry are high density, hardness,
melting temperature, elastic modulus, and conductivity in conjunction
with the low thermal expansion. The combination of these unique
properties explains the diverse applications of tungsten ranging from
home lighting to thermonuclear fusion first-wall
protection\cite{Yih:1979, Lassner:1999}.  With nanoscale tungsten
powders available at a reasonable cost, its usage will increase
greatly and a new approach is required to balance the size dependent
advantages against the temperature dependent limitations. Therefore,
it is of great importance to understand the thermal stability of
tungsten nanoparticles for their applications at higher temperatures.
It has been seen previously that nanoparticles exhibit significant
decrease in melting temperatures compared to infinite bulk
solids\cite{Borel:1981}.  This is related to the fact that the
liquid-vapor interface energy is generally lower than the average
solid-vapor interface energy\cite{Gulseren:1995}.  Based on
thermodynamics, a phenomenological relation between the melting
temperature and particle size has been obtained: the melting
temperature of a nanoparticle decreases as inversely proportional to
the particle diameter \cite{Pawlow:1909, Borel:1981, Gulseren:1995}.
It is also known that premelting, the phenomenon where the surface
atoms of nanoparticles lose their solid ordering and hence melt prior
to complete melting of the whole particle \cite{Lai:1996, Lai:1998,
  Pawlow:1909, Frenken, Ercolessi:1991, Lewis:1997}, plays an
important role in understanding the melting of nanoparticles.  On the
other hand, the Hall-Petch effect---the hardness increases in
proportional to the inverse square-root of grain size \cite{Hall:1951,
  Petch:1953}---suggests that significant opportunities exist if
nanoscale powders could be consolidated to full densities with
minimized coarsening. Hence, knowledge of accurate melting temperature
for different particle size plays an important role for the
advancement of present engineering and technological growth.

Molecular dynamics simulations offer an effective tool to study the
melting and coalescence of nanoparticles\cite{Shim:2002,
  Kim:1994:Fullerene}.  These atomistic simulations require accurate
atomic interaction potentials to compute the total energy of the
system.  First-principles calculations can provide the most reliable
interatomic potentials.  However, realistic simulations of the melting
of nanoparticles often require a number of atoms that renders these
methods impractical: they either require too much computer memory or
take too long to be completed in a reasonable amount of time.  One
alternative is to use empirical or semi-empirical interaction
potentials that can be evaluated efficiently.  In this study, we use
the modified embedded atom method (MEAM) originally proposed by Baskes
et al. \cite{Baskes:1989:MEAM, Baskes:1992:MEAM}.  MEAM was the first
semi-empirical atomic potential using a single formalism for fcc, bcc,
hcp, diamond-structured materials and even gaseous elements, in good
agreement with experiments or first-principles calculations
\cite{Baskes:1992:MEAM, Baskes:1994:MEAM-hcp}.  The MEAM is an
extension of the embedded-atom method (EAM) \cite{Daw:1989:EAM,
  Daw:1984:EAM, Daw:1983:EAM} to include angular forces.  Cherne et
al. made a careful comparison of MEAM and EAM calculations in a liquid
nickel system\cite{Cherne:2001:PRB-65}.

Atomistic simulations of a wide range of elements and alloys have been
performed using the MEAM potentials.  A realistic shear behavior for
silicon was first obtained using the MEAM by Baskes et al.
\cite{Baskes:1989:MEAM}.  The MEAM was also applied to various single
elements \cite{Baskes:1992:MEAM} and to silicon-nickel alloys and
interfaces \cite{Baskes:1994:MSMSE}.  Gall et al\cite{Gall:2000} used
the MEAM to model the tensile debonding of an aluminum-silicon
interface.  Lee and Baskes \cite{Lee:2000:MEAM} extended the MEAM to
include the second nearest-neighbor interactions. A new analytic
modified embedded-atom method (AMEAM) many-body potential was also
proposed and applied to several hcp metals, including
Mg\cite{Hu:2001:AMEAM, Hu:2003}.  For the Mg-Al alloy system, a set of
EAM potentials has been developed using the ``force matching'' method
by Liu et al\cite{Liu:1997:SURF}.  Recently, a new set of MEAM
potentials for Mg-Al alloy system was developed by Jelinek et
al\cite{Jelinek:2007}.  These new potentials show a significant
improvement over the previously published potentials, especially for
the surface formation, stacking faults, and point defect formation
energies.

The paper is organized in the following manner. In
Sec.~\ref{sec:Theory}, we give a brief review of the MEAM. In
Sec.~\ref{sec:Simulation}, the procedure for melting simulation is
presented.  MD simulation results are presented and discussed in
Sec.~\ref{sec:Results}.  Finally, in Sec.~\ref{sec:Conclusion}, we
summarize our findings.

\section{MEAM theory}
\label{sec:Theory}

The total energy $E$ of a system of atoms in the MEAM
\cite{Kim:2006:MEAM} is approximated as the sum of the atomic energies
\begin{equation}
  E = \sum_{i} E_i.
\end{equation}
The energy of atom $i$ consists of the embedding energy and the pair
potential terms:
\begin{equation}
  E_i = F_i(\bar\rho_{i}) + \frac{1}{2} \sum_{j \neq i}\phi_{ij}(r_{ij}).
\end{equation}
$F_i$ is the embedding function of atom $i$, $\bar\rho_{i}$ is the
background electron density at the site of atom $i$, and
$\phi_{ij}(r_{ij})$ is the pair potential between atoms $i$ and $j$
separated by a distance $r_{ij}$.  The embedding energy
$F_i(\bar\rho_{i})$ represents the energy cost to insert atom $i$ at a
site where the background electron density is $\bar\rho_{i}$. The
embedding energy is given in the form
\begin{equation}
  \label{eq:emb}
  F_i(\bar\rho_{i}) = A_{i} E_{i}^{0} \bar\rho_{i} \ln (\bar\rho_i),
\end{equation}
where the sublimation energy $E_i^0$ and parameter $A_i$ depend on the
element type of atom $i$.  The background electron density
$\bar\rho_i$ is given by
\begin{equation}
  \label{eq:rhobar}
  \bar\rho_{i} = \frac{\rho_{i}^{(0)}}{\rho_{i}^0} G(\Gamma_i),
\end{equation}
where
\begin{equation}
  \Gamma_i = \sum_{k=1}^3 \bar{t}_i^{(k)}
  \left(
    \frac{\rho_i^{(k)}}{\rho_i^{(0)}}
  \right)^2
\end{equation}
and
\begin{equation}
  G(\Gamma) = \sqrt{1 + \Gamma}.
\end{equation}
The zeroth and higher order densities, $\rho_i^{(0)}$, $\rho_i^{(1)}$,
$\rho_i^{(2)}$, and $\rho_i^{(3)}$ are given in
Eq.~(\ref{eq:part_den}).  The composition-dependent electron density
scaling $\rho_i^0$ is given by
\begin{equation}
  \rho_i^0 = \rho_{i0}Z_{i0} G( \Gamma_i^\text{ref} ),
\end{equation}
where $\rho_{i0}$ is an element-dependent density scaling, $Z_{i0}$
is the first nearest-neighbor coordination of the reference system, and
$\Gamma_i^\text{ref}$ is given by
\begin{equation}
  \Gamma_i^\text{ref} = \frac{1}{Z_{i0}^2}
  \sum_{k=1}^3 \bar{t}_i^{(k)} s_i^{(k)},
\end{equation}
where $s_i^{(k)}$ is the shape factor that depends on the reference
structure for atom $i$. Shape factors for various structures are
specified in the work of Baskes\cite{Baskes:1992:MEAM}.  The partial
electron densities are given by
\begin{subequations}
  \label{eq:part_den}
\begin{eqnarray}
  \label{eq:part_den_first}
  \rho_i^{(0)} & = & \sum_{j \neq i} \rho_j^{a(0)} S_{ij} \\
  \left( \rho_i^{(1)} \right)^2 & = &
  \sum_{\alpha}
  \left[
    \sum_{j \neq i} \rho_j^{a(1)} \frac{r_{ij\alpha}}{r_{ij}} S_{ij}
  \right]^2 \\
  \left( \rho_i^{(2)} \right)^2 & = &
  \sum_{\alpha, \beta}
  \left[
    \sum_{j \neq i} \rho_j^{a(2)} \frac{r_{ij\alpha}r_{ij\beta}}{r_{ij}^2} S_{ij}
  \right]^2 \nonumber \\
  & & -\frac{1}{3}
  \left[
    \sum_{j \neq i} \rho_j^{a(2)} S_{ij}
  \right]^2
  \\
  \left( \rho_i^{(3)} \right)^2 & = &
  \sum_{\alpha, \beta, \gamma}
  \left[
    \sum_{j \neq i} \rho_j^{a(3)}
    \frac{r_{ij\alpha}r_{ij\beta}r_{ij\gamma}}{r_{ij}^3} S_{ij}
  \right]^2 \nonumber\\
  & & -\frac{3}{5} \sum_{\alpha}
  \left[
    \sum_{j \neq i} \rho_j^{a(3)} \frac{r_{ij\alpha}}{r_{ij}} S_{ij}
  \right]^2,
  \label{eq:part_den_last}
\end{eqnarray}
\end{subequations}
where $r_{ij\alpha}$ is the $\alpha$ component of the displacement vector
from atom $i$ to atom $j$.  $S_{ij}$ is the screening function between
atoms $i$ and $j$ and is defined in Eqs.~(\ref{eq:scr}).  The atomic
electron densities are computed as
\begin{equation}
  \rho_i^{a(k)}(r_{ij}) =
  \rho_{i0} \exp
  \left[
    - \beta_i^{(k)} \left( \frac{r_{ij}}{r_i^0} - 1 \right)
  \right],
\end{equation}
where $r_i^0$ is the nearest-neighbor distance in the single-element
reference structure and $\beta_i^{(k)}$ is element-dependent
parameter.  Finally, the average weighting factors are given by
\begin{equation}
  \bar{t}_i^{(k)} = \frac{1}{\rho_i^{(0)}}
  \sum_{j \neq i} t_{j}^{(k)} \rho_j^{a(0)} S_{ij},
\end{equation}
where $t_{j}^{(k)}$ is an element-dependent parameter.

The pair potential is given by
\begin{align}
  \label{eq:pair}
  \phi_{ij}(r_{ij}) &= \bar\phi_{ij}(r_{ij}) S_{ij} \\
  \begin{split}
  \bar\phi_{ij}(r_{ij}) &= \frac{1}{Z_{ij}}
  \left[ 2E_{ij}^u (r_{ij}) 
    -F_i\left(\frac{Z_{ij}\rho^{(0)}_j(r_{ij})}{Z_i\rho^0_i}\right) \right. \\
    &\quad \left. -F_j\left(\frac{Z_{ij}\rho^{(0)}_i(r_{ij})}
        {Z_j\rho^0_j}\right) \right] \label{eq:rhohat}
\end{split} \\
  E_{ij}^u(r_{ij}) &= -E^0_{ij} \left( 1 + a_{ij}^*(r_{ij})\right)
  e^{-a_{ij}^{*}(r_{ij})} \\
  a_{ij}^{*} &= \alpha_{ij} \left( \frac{r_{ij}}{r_{ij}^0} - 1 \right),
\end{align}
where $\alpha_{ij}$ is an element-dependent parameter.  The
sublimation energy $E^0_{ij}$, the equilibrium nearest-neighbor
distance $r_{ij}^0$, and the number of nearest-neighbors $Z_{ij}$ are
obtained from the reference structure. 

The screening function $S_{ij}$ is designed so that $S_{ij} = 1$ if
atoms $i$ and $j$ are unscreened and within the cutoff radius $r_c$,
and $S_{ij} = 0$ if they are completely screened or outside the cutoff
radius. It varies smoothly between 0 and 1 for partial screening. The
total screening function is the product of a radial cutoff function
and three-body terms involving all other atoms in the system:
\begin{subequations}
  \label{eq:scr}
  \begin{align}
    \label{eq:scr_first}
    S_{ij} &= \bar{S}_{ij} f_c \left( \frac{r_c - r_{ij}}{\Delta r}
    \right) \\
    \bar{S}_{ij} &= \prod_{k\ne i,j}S_{ikj} \\
    S_{ikj} &= f_c \left(\frac{C_{ikj} - C_{\text{min},ikj}}
      {C_{\text{max},ikj} - C_{\text{min},ikj}} \right) \\
    C_{ikj} &= 1 + 2 \frac{r_{ij}^2 r_{ik}^2 + r_{ij}^2 r_{jk}^2 
      - r_{ij}^4}{r_{ij}^4 - \left( r_{ik}^2 - r_{jk}^2 \right)^2} \\
    f_c\left(x\right) &=
    \begin{cases}
      1 & x \geq 1\\
      \left[ 1 - \left( 1 - x )^4 \right) \right]^2 & 0<x<1 \\
      0 & x \leq 0\\
    \end{cases}
    \label{eq:scr_last}
  \end{align}
\end{subequations}
Note that $C_{\text{min}}$ and $C_{\text{max}}$ can be defined
separately for each $i$-$j$-$k$ triplet, based on their element
types. The parameter $\Delta r$ controls the distance over which the
radial cutoff function changes from 1 to 0 near $r=r_c$.

\section{Molecular dynamics simulation}
\label{sec:Simulation}

\subsection{Atomic potential}

We use the MEAM potential parameters for tungsten (W) proposed by
Baskes \cite{Baskes:1992:MEAM}.  The potential parameters that are used for
our simulation of W nanoparticles are listed in
Table~\ref{tab:Parameters}.  These parameters are obtained by fitting
the room temperature elastic properties using bcc as the reference
structure. $C_{\text{max}}$ and $C_{\text{min}}$ are chosen to
consider only the first nearest-neighbor interactions
\cite{Baskes:1997:MEAM}.

We validate the potential by computing different physical properties
of tungsten systems and comparing them with DFT calculations.  The
results are compared with those of DFT calculations as shown in
Table~\ref{tab:properties}.  Energy calculations and geometry
optimizations of various structures were performed using Bl\"{o}chl's
all-electron projector augmented wave (PAW) method\cite{Blochl:1994}
as implemented by Kresse and Joubert
\cite{Kresse:1999:PhysRevB.59.1758}.  For the treatment of electron
exchange and correlation, we use the generalized gradient
approximation (GGA) using Perdew-Burke-Ernzerhof scheme
\cite{Perdew:1996}.

\subsection{Simulation Procedure}

We performed a detailed MD simulation of the melting of unsupported
spherical bcc W nanoparticles, 2--12 nm in diameter (259--56905
atoms).  The surface boundary condition was free and no external
pressure was applied.  Each nanoparticle was constructed by cutting
out atoms within a specified radius from the tungsten bulk in bcc
structure.  The equations of motion were integrated using time steps
$\Delta t = 4\times 10^{-15}$ s.  We begin each MD run by randomizing
the atomic velocities of the nanoparticle according to the
Maxwell-Boltzmann distribution.  We increase the temperature of the
heat bath in steps of $\Delta T = 100$~K from the initial temperature
$T_i = 500$~K to the final temperature up to $T_f = 4000$~K.  We let
the nanoparticles come to equilibration for 50 000 time steps at each
temperature.  Statistical (time-averaged) data for the energetics are
collected after the system has adjusted to the new temperature, which
is typically after 25 000 time steps following a temperature increase.
For the particles of diameters less than 8~nm, 20 000 time steps were
used to adjust the particles to each new temperature.  The isothermal
condition was maintained by using Nos\'{e}-Hoover thermostat
\cite{Hoover:1985, Nose:1984}.

\section{Results and Discussion}
\label{sec:Results}

\begin{figure}
  \epsfxsize=8cm
  \begin{center}
    \mbox{\epsfbox{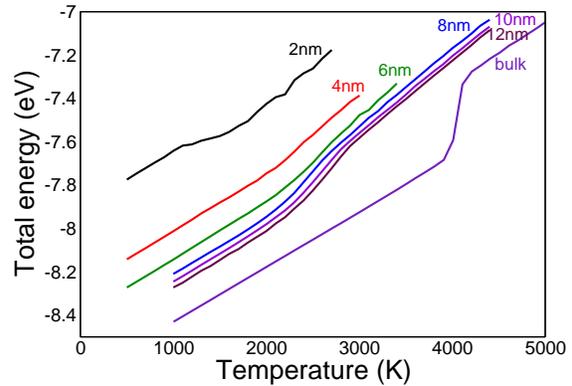}}
  \end{center}
  \caption{Internal energies of the W nanoparticles with different
    diameters as a function of temperature. The same data for W bulk
    are also shown.}
  \label{fig:energy-temp}
\end{figure}

The most straightforward method to identify the melting of atomistic
structures in MD simulations is to monitor the variation of the
internal energy with temperature. Fig.~\ref{fig:energy-temp} shows the
internal energies of the W nanoparticles with different diameters as a
function of temperature.  It is clearly seen from the
Fig.~\ref{fig:energy-temp} that each internal energy curve goes from
one linear region to another.  The overall melting is clearly
identified by the abrupt ``jump'' in the internal energy curve.  the
height of the jump is a measure of $\Delta H_m$, the amount of heat
required for melting, and it decreases as the size of nanoparticle
decreases.  The melting temperatures calculated based on the present
MD simulation of W nanoparticles are listed in
Table~\ref{tab:Tm-diameter}.  We note that the melting temperature of
bulk W from our MD simulation, 3900~K, is slightly higher than the
experimentally measured value of 3695~K \cite{Emsley:1998}.  The
discrepancy in this result is mainly due to the super-heating of the
simulated lattice, as it has been observed that the confined lattice
without free surface can be significantly superheated \cite{Jin:1999,
  Lu:1998}.  Although, it is not the main focus of this study, one can
follow the procedure prescribed by Morris et al \cite{Morris:1994} to
establish co-existence of solid and liquid phases to determine the
melting temperature of the bulk W system without super-heating.  More
importantly, we also note that the melting temperature decreases
drastically as the size of the particle decreases. This result
suggests that the thermal stability of small nanoparticles must be
carefully investigated before they can be used in applications such as
nano-devices.

The melting behavior of 2~nm particle seems to be different from those
of bigger particles: at the onset of the melting, the internal energy
curve dips down before climbing up again.  A similar behavior has been
observed in the melting of small Au nanoparticles of diameters up to
2.8~nm \cite{Shim:2002}.  The onset of melting provides surface atoms
an opportunity to rearrange themselves to optimize the local
morphology and lower their portion of the internal energy. For
extremely small particles, where the surface area to volume ratio is
large, this will cause the total internal energy of the particle to
decrease briefly as shown in Fig.~\ref{fig:energy-temp}. However, a
further detailed study focusing on small nanoparticles will be
required to elucidate this peculiar behavior.

\begin{figure}
  \epsfxsize=8cm
  \begin{center}
    \mbox{\epsfbox{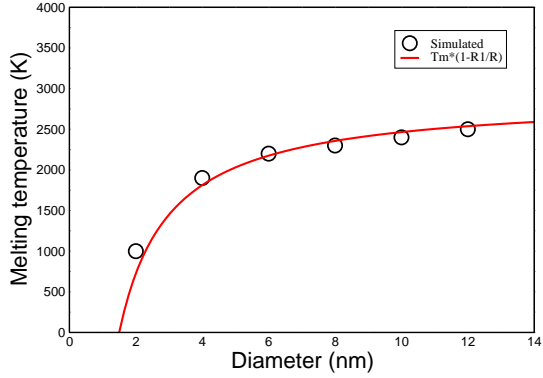}}
  \end{center}
  \caption{Size dependence of the melting
    temperatures of W nanoparticles.  Symbols represent the calculated
    values from the present MD simulation and the solid line is
    calculated in terms of Eq.~(\ref{eq:Tm-Pawlow}).}
  \label{fig:Tm-diameter}
\end{figure}

The variation of the melting temperature with the size of the W
nanoparticles is plotted in Fig.~\ref{fig:Tm-diameter}.  The melting
point depression of W nanoparticles exhibit the same qualitative
behavior found in the MD simulation of Au nanoparticles
\cite{Ercolessi:1991, Shim:2002}.  A similar size dependence of
melting point depression has been observed experimentally over a broad
range of particle sizes for particles in cluster beams
\cite{Bertsch:1997, Berry:1990, Schmidt:1998, Schmidt:1997} as well as
particles on substrates \cite{Allen:1986, Peters:1998, Lai:1996,
  Lai:1998}.

For spherical particles of diameter $R$, a melting temperature
$T_m(R)$ can be obtained phenomenologically
\cite{Gulseren:1995, Buffat:1976, Pawlow:1909} by equating the Gibbs
free energies of solid and liquid spherical clusters, assuming
constant pressure conditions:
\begin{equation}
  \label{eq:Tm-Pawlow}
  T_{m}(R) = T_{m}^{\star} \left( 1 - \frac{R_1}{R} \right),
\end{equation}
where $T_{m}^{\star}$ is the melting temperature of the bulk tungsten
and $R_1$ is a parameter related to physical quantities such as the
solid and liquid densities, the bulk latent heat of melting, and
solid-vapor and liquid-vapor interface energies.  In obtaining this
model, the surface energy anisotropy of the solid is not taken into
account, and the possibility of inhomogeneous phases (such as a liquid
layer due to premelting) is also neglected.  The solid line in
Fig.~\ref{fig:Tm-diameter} corresponds to the simple thermodynamical
model of Eq.~(\ref{eq:Tm-Pawlow}), with constant parameters
$T^{\star}$ = 2900~K and $R_1$ = 1.5~nm.  The curve shows clearly that
the melting point of W nanoparticles decrease according to $1/R$
dependence as predicted in Eq.~(\ref{eq:Tm-Pawlow}).  However, the
fitted value of $T^{\star}$ is significantly lower than the melting
temperature of the bulk tungsten.  This result indicates that the
characteristics of the curve is likely to change for nanoparticles
with larger diameters, and further study with larger nanoparticles
will be needed to test the applicability of this model to W
nanoparticles.

Hanszen \cite{Hanszen:1960} proposed another model of melting in terms
of classical thermodynamics assuming that a liquid over-layer forms at
the surface of the solid cluster and grows towards the solid core,
below the melting point \cite{Efremov:2000, Zhang:2000}.  When the
liquid layer thickness exceeds a critical thickness, the whole cluster
melts homogeneously.  In this model, the melting point $T_m(R)$ of W
nanoparticles with diameter $R$ can be expressed as
\cite{Lai:1996, Lai:1998}
\begin{equation}
  \label{eq:Tm-Henszen}
  T_{m}(R) = T_{m}^{\star} \left( 1-\frac{R_1}{R-t_0}+\frac{R_2}{R} \right),
\end{equation}
When the data of Table~\ref{tab:Tm-diameter} were fitted to
Eq.~(\ref{eq:Tm-Henszen}), we obtained negligibly small values for
$t_0$ and $R_2$, thus returning to the model of
Eq.~(\ref{eq:Tm-Pawlow}).

\begin{figure}
  \epsfxsize=8cm
  \begin{center}
    \mbox{\epsfbox{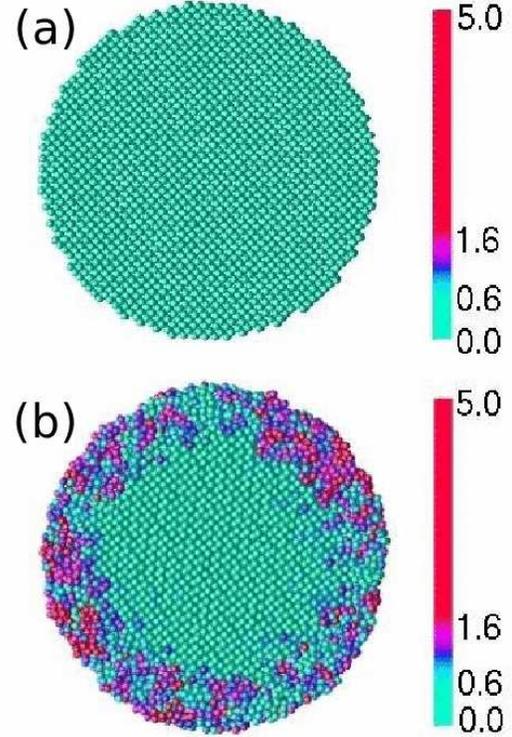}}
  \end{center}
  \caption{(color online). The cross section of W
    nanoparticle with a diameter 10~nm through the center of the
    particle showing the displacement vectors in the interval of 6 ps
    at the temperature of (a) 300~K and (b) 2000~K.  The color and the
    size of the spheres represent the magnitude of the displacement
    vectors.}
  \label{fig:displacement}
\end{figure}

Fig.~\ref{fig:displacement} shows the cross sections of a W
nanoparticle with a diameter 10~nm through the center of the particle.
The displacement vectors at different times during the MD simulation
run at the temperature 2000~K are plotted.
Fig.~\ref{fig:displacement} shows that at a temperature below the
melting point the atoms in the entire nanoparticle vibrate in their
places while retaining their bcc crystal structure. As the temperature
increase, several layers of atoms start to lose their periodicity and
form a liquid shell as shown in Fig.~\ref{fig:displacement}(b).  Once
the thickness of the liquid layer reaches a critical thickness, the
whole nanoparticle melts.  Our MD simulation confirms the experimental
observation that nanoscale materials simultaneously display solid-like
and liquid-like characteristics, and under thermodynamic equilibrium,
a fraction of the atoms in the outer shell of the particle exhibit
liquid-like behavior and the remaining fraction in the inner core act
as solid \cite{Lai:1998}.  Hence, melting point depression and the
presence of disorder in nanoscale W powders will play an important
role in various industries, including microelectronic industries such
as printed circuit board drill bits \cite{Gille:2002}.

\begin{figure}
  \epsfxsize=8cm
  \begin{center}
    \mbox{\epsfbox{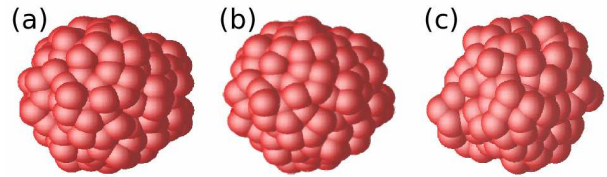}}
  \end{center}
  \caption{Snapshots of MD simulation at (a) 300~K, (b) 900~K, and (c)
    1500~K (above the melting temperature) for 2~nm particle.}
  \label{fig:snapshots}
\end{figure}

Fig.~\ref{fig:snapshots} shows a few snapshots of MD simulation for a
small W nanoparticle with the diameter of 2~nm.  We note that our
simulations does not show pronounced faceted or step-like structures.
We found a similar result when the nanoparticles are heated to the
melting temperature and cooled down slowly.  Our results are in good
agreement with an earlier experiment that found no evidence for a
faceted or step-like microstructure in a single tungsten crystal
\cite{Martin:1939}.

\section{Conclusions}
\label{sec:Conclusion}

The thermal stability of unsupported W nanoparticles has been
investigated using a MD simulation. The MEAM potential was used to
described the interatomic interactions.  W nanoparticles melt at a
temperature that is lower than the bulk melting temperature.  The
result of our present calculation shows that the melting temperature
to be approximately a decreasing function of inverse radius. We found
that W nanoparticle melting is preceded by surface melting effects of
its outer skin, similar to the melting of spherical clusters of many
other elements.

\section{Acknowledgment}

The authors are grateful to the Center for Advanced Vehicular Systems
at Mississippi State University for supporting this study.  Computer
time allocation has been provided by the High Performance Computing
Collaboratory (HPC$^2$) at Mississippi State University.  

\bibliographystyle{unsrt}
\bibliography{nanopowder,DFT,MEAM}


\newpage

\begin{table*}[ht]
  \caption{The MEAM potential parameters 
    for W from Ref.~\citen{Baskes:1997:MEAM}. 
    $E^0$ is the sublimation energy, 
    $r^0$ is the equilibrium nearest-neighbor distance, 
    $A$ is the scaling factor for the embedding energy,
    $\alpha$ is the exponential decay factor for the universal energy
    function,
    $\beta^{(0-3)}$ are the exponential decay factors for the atomic
    densities, $t^{(0-3)}$ are the weighting factors for the atomic
    densities, $C_{\text{max}}$ and $C_{\text{min}}$ are the screening
    parameters.}
  \label{tab:Parameters}  
  \begin{center}
    \begin{tabular}{cccccccccccccc}
      \hline
      $E^0$ [eV] & $r^0$ [\AA] & $A$ & $\alpha$ &
      $\beta^{(0)}$ & $\beta^{(1)}$ & $\beta^{(2)}$ & $\beta^{(3)}$ &
      $t^{(0)}$ & $t^{(1)}$ & $t^{(2)}$ & $t^{(3)}$ &
      $C_{\text{max}}$ & $C_{\text{min}}$ \\
      \hline
      8.66 & 2.74 &  0.98 &  5.63 &
      3.98 & 1.00  &  1.00 &  1.00 &
      1.00 & 3.16  &  8.25 & -2.70 &
      2.8  & 2.0   \\
      \hline
    \end{tabular}
  \end{center}
\end{table*}

\begin{table}[!tbp]
  \caption{Calculated physical properties of W
    using the present MEAM parameters in comparison with DFT
    calculations. $B_{\text{0}}$ is the bulk modulus (GPa); 
    $C_{\text{11}}$, $C_{\text{12}}$, $C_{\text{44}}$
    are the elastic constants (GPa); $E_{\text{(100)}}$, $E_{\text{(110)}}$,
    $E_{\text{(111)}}$ are surface energies of corresponding surfaces 
    (mJ/m$^2$); $\Delta E$'s are the structural energy differences (eV/atom).
  }
  \label{tab:properties} 
  \begin{center}
    \begin{tabular}{ccr}
      \hline
      Parameter & DFT & MEAM \\
      \hline
      $B_{\text{0}}$ & 330 & 270 \\
      $(C_{\text{11}}-C_{\text{12}})/2$  & 190  & 160 \\
      $C_{\text{44}}$ & 280  & 160 \\
      $E_{\text{(100)}}$ & 7810 &  5980 \\
      $E_{\text{(110)}}$ & 6390 &  5660 \\
      $E_{\text{(111)}}$ & 7190 &  5030 \\
      $\Delta E_{\text{bcc} \to \text{fcc}}$ & 0.494 & 0.325 \\
      $\Delta E_{\text{bcc} \to \text{hcp}}$ & 0.397 & 2.168 \\
      \hline
    \end{tabular}
  \end{center}
\end{table}

\begin{table}
  \caption{Melting temperatures of W 
    nanoparticles with different diameters}
  \label{tab:Tm-diameter}
  \begin{center}
    \begin{tabular}{rrr}
      \hline
      Diameter (nm) & No. of atoms & $T_m$ (K) \\
      \hline
      2.0  & 259    & 1000 \\
      4.0  & 2085   & 1900 \\
      6.0  & 7119   & 2200 \\
      8.0  & 16865  & 2300 \\
      10.0 & 33079  & 2400 \\
      12.0 & 56905  & 2500 \\
      Bulk & ${\infty}$ & 3900 \\
      \hline
    \end{tabular}
  \end{center}
\end{table}

\end{document}